\documentclass[epj,nopacs]{svjour}
\usepackage{amsmath}
\usepackage{amssymb}
\usepackage{latexsym}
\usepackage[dvips]{graphicx}
\usepackage[english]{babel}

\def\bge{\begin{equation}}
\def\ene{\end{equation}}
\def\bgea{\begin{eqnarray}}
\def\enea{\end{eqnarray}}

\def\bge{\begin{equation}}
\def\ene{\end{equation}}
\def\bgea{\begin{eqnarray}}
\def\enea{\end{eqnarray}}

\def\ls{\raise 1.5pt\hbox{$\,<\;$}\kern -10.5pt\lower3.5pt
          \hbox{$\sim$}\kern 1.5pt} %%% less or similar
\def\gs{\raise 1.5pt\hbox{$\,>\,$}\kern -9.5pt\lower3.5pt
          \hbox{$\sim$}\kern 1.5pt} %%% greater or similar

\usepackage{color}

\usepackage{ulem} % use: \sout{Striked out text}

\begin{document}
\sloppy
\title{Thermodynamics of the $R_{\rm h}=ct$ Universe: \\
A Simplification of Cosmic Entropy}
\author{Fulvio Melia\thanks{John Woodruff Simpson Fellow.}}
\institute{Department of Physics, the Applied Math Program, and Department of Astronomy, \\
              The University of Arizona, Tucson, AZ 85721,
              \email{fmelia@email.arizona.edu}}

\authorrunning{Melia}
\titlerunning{Cosmic Entropy}

\date{\today}
%\date{March 10, 2010}
%\date{May 07, 2010}
%\date{August 8, 2018}
%\date{}

\abstract{In the standard model of cosmology, the Universe began its expansion with an
anomalously low entropy, which then grew dramatically to much larger values consistent
with the physical conditions at decoupling, roughly 380,000 years after the Big Bang.
There does not appear to be a viable explanation for this `unnatural' history, other than
via the {\it generalized} second law of thermodynamics (GSL), in which the entropy of the bulk,
$S_{\rm bulk}$, is combined with the entropy of the apparent (or gravitational) horizon,
$S_{\rm h}$. This is not completely satisfactory either, however, since this approach
seems to require an inexplicable equilibrium between the bulk and horizon temperatures.
In this paper, we explore the thermodynamics of an alternative cosmology known as the
$R_{\rm h}=ct$ universe, which has thus far been highly successful in resolving many
other problems or inconsistencies in $\Lambda$CDM. We find that $S_{\rm bulk}$ is
constant in this model, eliminating the so-called initial entropy problem simply and
elegantly. The GSL may still be relevant, however, principally in selecting the arrow
of time, given that $S_{\rm h}\propto t^2$ in this model.}
%    \PACS{{04.20.Ex},\  {95.36.+x},\  {98.80.-k},\  {98.80.Jk}}
\maketitle

% ---------------------------------------------------------------------------------
\section{Introduction}
A conjecture known as the `past hypothesis' posits that the entropy of the observable Universe
is increasing monotonically, and must therefore have been lower---significantly lower---at
earlier times \cite{Layzer:1975,Price:1996,Albert:2000,Earman:2006}. But this position
appears to be at odds with the observed cosmic microwave background (CMB) which, in the
context of standard $\Lambda$CDM, suggests that the Universe was close to thermal and
chemical equilibrium, i.e., in a state of very high entropy, a mere $\sim 380,000$
years after the Big Bang. These two conflicting views give rise to what is commonly
referred to as the `cosmic initial entropy problem' (IEP) \cite{Carroll:2004,Patel:2017}.

The Universe seems to be homogeneous on scales exceeding several hundred Mpc, so one may
reasonably assume that no entropy is flowing between neighboring volumes across distances
larger than this. And since physical processes, such as stellar evolution and
black-hole accretion, appear (on balance) to be increasing the cosmic entropy locally, it is
difficult to understand why we would be living in a portion of the Universe with anomalously
low entropy today \cite{Zemansky:1997,Frautschi:1982,Albrecht:2002,Penrose:2004,Egan:2010},
if its entropy shortly after the Big Bang was as high as it appears to have been at
decoupling (i.e., $z_{\rm dec}\sim 1080$).

The problem arises because blackbody radiation in equilibrium contains the largest amount
of entropy, with a volume density
\begin{equation}
s_{\rm BB}={4\over 3}a_{\rm rad}T_{\rm BB}^3\;,\label{eq:CMBentropydensity}
\end{equation}
in terms of the blackbody temperature $T_{\rm BB}$ and radiation constant $a_{\rm rad}$. As
the Universe expands, $T_{\rm BB}$ scales as $(1+z)\propto a(t)^{-1}$, where $a(t)$ is the
expansion factor in the Friedmann-Lema\^{i}tre-Robertson-Walker (FLRW) metric (Eq.~\ref{eq:FLRW}).
At the same time, any given proper volume $V$ scales as $a(t)^3$, so the blackbody entropy
$S_{\rm BB}=s_{\rm BB}V$ must have remained constant \cite{Frautschi:1982}. Then how
could the CMB have been created at $z_{\rm dec}$ with the same entropy it has today, when
all the other indications are that the total entropy should have been much smaller back then?
In the context of $\Lambda$CDM, solutions to the IEP must therefore simultaneously explain
why the Universe was initially in a very low entropy state (as is seemingly required by the
second law), and why the observed CMB acquired such high entropy. To be clear, a very low
initial entropy on its own may not be a problem if, e.g., the Universe was created from
`nothing' and has evolved away from that {\it a priori} unique initial state to which it would
never return \cite{Vilenkin:1982,Linde:1984,Hartle:1983}.

This basic picture has undergone several refinements over the past two decades, and we shall
summarize the key steps taken to better understand the origin of cosmic entropy in the next
section. The principal goal of this paper, however, is to examine the IEP and related entropy
issues in the context of an alternative FLRW cosmology known as the $R_{\rm h}=ct$ universe,
which is receiving support from many observational tests (see, e.g., Table~2 of
ref.~\cite{Melia:2018a}), and several compelling theoretical arguments, notably the Local
Flatness Theorem in general relativity \cite{Melia:2019a}. A complete description of this
model and its observational and theoretical foundation may be found in ref.~\cite{Melia:2020a}.
In \S~2, we shall provide a more extended background of the IEP, and then summarize the
inventory of cosmic entropy---as we know it today---in \S~3. We shall discuss the
simplification of cosmic entropy when viewed from the perspective of $R_{\rm h}=ct$ in
\S~4, and then conclude in \S~5.

\section{Background}
\subsection{The Basic Picture}
Today, serious attempts at resolving the IEP tend to rely on the inclusion of a hypothesized
gravitational entropy \cite{Penrose:2004,Kolb:1994} which, at first, may seem
counterintuitive. A concentration of particles in a kinetically-dominated system tend to
diffuse until they reach a homogeneous distribution, corresponding to thermal equilibrium
and maximum entropy. A smooth distribution in a gravitationally-dominated system, on the
other hand, corresponds to low gravitational entropy, which increases as initially small
clumps collapse gravitationally to eventually form black holes \cite{Bekenstein:1973}. (The
caveat here, however, is that this simple-minded consideration may be missing other
contributions to gravitational entropy, as we shall discuss in \S~3.A below.) Indeed,
we shall see in \S~3 that supermassive black holes are among the largest contributors
to the bulk entropy of the Universe \cite{Egan:2010}. A simple solution to the IEP may
therefore be that the Universe began with a very low entropy at the Big Bang,
which increased rapidly by $z_{\rm dec}$ to produce the observed CMB, and then
continued to increase to values {\it even larger} than $S_{\rm BB}$ today, generated principally
by the formation of supermassive black holes at the centers of most galaxies. It has been postulated
that this gravitational entropy may increase even further if black holes eventually evaporate
due to Hawking radiation \cite{Hawking:1983,Zurek:1982,Page:2005}.

But the thermodynamics of gravitational fields remains contentious
\cite{Binney:2008,Wallace:2010,Penrose:2004} (see also \S~3.A), and black-hole entropy does
not explain why the Universe would have been in a very low entropy state to begin with. Instead,
the inflationary paradigm \cite{Guth:1981,Linde:1982,Albrecht:1982} is often invoked to explain why
the initial entropy of the Universe might have been so low \cite{Carroll:2004,Davies:1983,Albrecht:2009},
based on the notion that its value was established during the slow-roll evolution of the
inflaton potential. Unfortunately, this idea has its own detractors, given that it contains
a hidden fine-tuning of initial conditions implicit in the assumption of a pre-inflationary
patch with exceedingly low entropy, producing an `unnatural' and unlikely state for the inflaton
field $\phi$ \cite{Page:1983,Penrose:1989}. It has been noted, e.g., that it requires much less
fine-tuning for the Universe to have been put in some state that evolved into the present conditions
than to have undergone an early period of inflation \cite{Carroll:2004}.

Worse, as the precision of cosmological measurements continues to improve, the argument against
such an `unnatural' beginning continues to gain support. A recent study using the latest
{\it Planck} data release \cite{Planck:2018} suggests that the primordial power spectrum
${\mathcal{P}}(k)$ has a hard cutoff, $k_{\rm min}=(3.14\pm 0.36)\times 10^{-4}\;{\rm Mpc}^{-1}$
\cite{MeliaLopez:2018}. A zero value of $k_{\rm min}$ is therefore excluded at a confidence
level exceeding $\sim 8\sigma$. But this is not what the slow-roll inflationary paradigm was
counting on. In order to simultaneously solve the horizon problem and generate a near scale-free
fluctuation spectrum consistent with the CMB observations, ${\mathcal{P}}(k)$ should have
extended well below the measured $k_{\rm min}$ \cite{LiuMelia:2020}. As a result, most (if not
all) slow-roll inflationary models proposed thus far, fail to accommodate this minimum cutoff.
Even extensions to the basic picture, incorporating kinetic-dominated or radiation-dominated
pre-inflationary phases, have no impact on this conclusion. It therefore appears very unlikely
that the very low entropy in the early $\Lambda$CDM Universe could have been due to inflation
as we know it today.

An alternative explanation for the very low initial entropy has been around much longer, and
began with Boltzmann himself \cite{Boltzmann:1895,Boltzmann:1897}, who proposed that the
low-entropy Universe we live in started as a random fluctuation in an otherwise maximal
entropy state. This is a highly unlikely event, of course, and the probability of it
happening drops sharply as the size of the fluctuation increases but, given an infinite
timeline, it is bound to happen sooner or later. This idea of an equilibrium state
permitting a low entropy fluctuation is an accepted notion of why the $\Lambda$CDM
Universe might have begun its expansion with an incredibly low entropy, but detractors
question the size of the Universe in such a state.

Coupled with an anthropic hypothesis, the fluctuation theorem \cite{Evans:1994} allows one to
quantify the relative probabilities of us living in a Universe of various sizes, showing that
small low-entropy universes are exponentially more likely than large ones. But this is where
the Boltzmann hypothesis breaks down. The Universe does not need to be this big for sentient
beings to exist within it. We see trillions of stellar systems, many more than are necessary
for life to emerge (we believe), so the anthropic principle cannot be a factor in this
argument \cite{Carter:1973,Barrow:1988}. It appears that the Boltzmann equilibrium
hypothesis is no more `natural' than the initial conditions required for inflation to
work.

It is fair to say that the IEP remains contentious for various reasons and is largely
unsolved---at least without the inclusion of an exploratory new feature having to do
with the entropy of an apparent horizon (see \S~3.B below). Meanwhile, gravitational
entropy might explain why the CMB had such large entropy at $z_{\rm dec}$, even though
$S_{\rm BB}$ has remained constant, if the gravitational assembly of large structures
and supermassive black holes continued to increase the Universe's entropy since then,
which would thus be much greater than that of the blackbody radiation on its own. But
neither the equilibrium models nor the inflationary paradigm can successfully account
for the very low initial entropy without some criticism, much of it driven by our aversion
towards a lack of `naturalness.' If initial conditions arose randomly, with a uniform
probability of microstates, we would expect our Universe to have been born in a state
of maximum entropy (if the number of degrees of freedom is finite---more on this below),
representing thermal equilibrium, not the exceedingly unlikely low-entropy fluctuation we
appear to be in according to $\Lambda$CDM.

\subsection{A More Recent Refinement}
We have made no reference thus far to the issue of what is actually observable but, in
retrospect, it makes little sense to discuss the evolution of entropy without invoking
the observer who makes the measurements. The Universe may be infinite (which is suggested
by the apparent spatial flatness inferred from cosmological observations), but each observer
can `see' only a finite volume within it. It is reasonable to assume, therefore, that the
application of the first and second laws to cosmology ought to be physically meaningful only
when talking about measurements made consistently by the same observer within their
observable patch of the Universe.

Two developments in black-hole physics bolster this argument quite compellingly: black-hole
complementarity \cite{Susskind:1993} and the holographic principle
\cite{Hooft:1993,Susskind:1995,Bousso:2002}. The former postulates that consistent (and, if
necessary, complementary) descriptions of physical situations can only include events within
the `horizon' of a single observer. In the case of black holes, the identification of a horizon
within their static or stationary spacetimes is unambiguous---it is uniquely the {\it event} horizon.
But the situation has not been so clear in the cosmic setting, as we shall see shortly. Nevertheless,
the black-hole example does suggest that an analogous horizon must be utilized in cosmology
as well. (We refer the reader to the Appendix for a detailed description and comparison
of the apparent, particle and event horizons.)

One of the more important consequences of the holographic principle is the covariant
entropy bound \cite{Bousso:1999}, which limits the entropy contained within any given, finite
region. Ultimately, this limit is related to the area of a null bounding surface measured
in Planck units. And since the area of an appropriately defined horizon is finite, the
causally-connected volume within it can only contain a finite number of degrees of freedom.
As noted earlier, this makes it very unlikely for the early Universe to have fluctuated
into an anomalously low-entropy state, but it opens up the possibility of `generalizing'
the second law by including a contribution---perhaps even a dominant contribution---from the
horizon itself \cite{Zurek:1985,Frolov:1993,Sorkin:1997,Flanagan:2000}.

As we shall see in \S~3.B below, however, identifying the appropriate cosmological horizon to
use in such a generalization has not been straightforward. Experience with black-hole horizons
reasonably suggested at first that one should use the cosmic `event' horizon. This idea certainly
seemed to work in de Sitter space, but it hasn't worked in other cosmologies. Some cosmological
models don't even have an event horizon. Trial and error eventually revealed that the most likely
bounding surface to employ in a generalized second law is actually the gravitational, or `apparent'
horizon (see ref.~\cite{Melia:2018b} for a pedagogical description). As we now understand it, the
event horizon works for de Sitter (and presumably also for black holes) only because its spacetime
is static, and the event horizon therefore coincides with the apparent horizon in that model
(see the Appendix). We shall return to this feature of the entropy problem shortly, but first we
consider how much entropy the Universe actually possesses, and what its dominant contributions
appear to be.

\section{Inventory of Cosmic Entropy}
\subsection{Cosmic entropy in the Bulk}
Irreversible processes are apparently increasing at least some contributions to the cosmic entropy
on all scales, from gravitational clustering in the largest volumes, to dissipative accretion through
disks, fusion in the interior of stars and their explosive deaths, all the way down to planetary
activity in the form of weather, chemical and biological processes
\cite{Frautschi:1982,Lineweaver:2008}.

The present cosmic entropy budget has been estimated by various workers
\cite{Kolb:1981,Frautschi:1982,Penrose:2004,Bousso:2007,Frampton:2008,Frampton:2009,Egan:2010}
over the past several decades, and we shall here largely adopt their key results.  As we
go through these contributions, however, it is important to keep in mind that most of
these estimates were based on a poorly defined measure of the `observable' universe,
which has undergone significant theoretical evolution with the continued refinement
of how one should define what is actually `observable' (see \S~2.B above and \S~3.B
below). In the Appendix, we discuss what has typically been assumed in the past, and
why the so-called `particle' horizon is not a correct  measure of the proper size of
the observable Universe. Neither is the event horizon; indeed, some cosmological
models don't even possess one. The correct proper distance \cite{Melia:2013a,Melia:2018b}
representing what an observer actually sees is $R_{\gamma,\,{\rm max}}\sim \eta R_{\rm h}$,
where $R_{\rm h}\equiv c/H$ is the radius of the gravitational (or apparent) horizon,
written in terms of the redshift-dependent Hubble parameter $H(z)$. The constant $\eta$
is cosmology-dependent, but tends to be $\approx 1/2$ in the majority of cases. The
notable exception is de Sitter space, for which $\eta=1$.

\begin{table*}
\centering
{\footnotesize
 \renewcommand{\arraystretch}{0.7}
 \caption{Entropy of the Observable Universe.}
 \begin{tabular}{lrr}
 & & \\
 \hline\hline
 & & \\
 Contribution & Entropy Density $s$ &  Entropy $S$ \qquad\, \\
              & $\;\;(k_{\rm B}$ m$^{-3})$ \qquad\qquad  & ($k_{\rm B}$) \qquad\quad\;\; \\
 & & \\
 \hline
 & & \\
 SMBHs & $(3.7-16.6)\times 10^{23}$ & $(2.6-11.5)\times 10^{101}$ \\
 & & \\
 Stellar BHs & & \\
 ($\lesssim 15\;M_\odot$) & $(1-64)\times 10^{16}$ & \qquad $(0.7-44.2)\times 10^{94}$ \\
 & & \\
 CMB $(S_{\rm CMB})$  & $(1.478\pm0.003)\times 10^9$ & $(1.02\pm 0.02)\times 10^{87}$ \\
 & & \\
 Other & $\lesssim 10^9$ & $\lesssim 6.9\times 10^{86}$ \\
 & & \\
 \hline
 & & \\
 Total $(S_{\rm bulk})$ & $(3.7-16.6)\times 10^{23}$ & $(2.6-11.5)\times 10^{101}$ \\
 & & \\
 \hline
 & & \\
 Speculative & & \\
 Massive Halo & & \\
 BHs ($\lesssim 10^5\;M_\odot$) & $\sim 10^{25}$ & $\sim (6.9\pm0.1)\times 10^{102}$ \\
 & & \\
 \hline\hline
 & & \\
 \multicolumn{3}{c}{
 \vbox{\hbox{{\it Notes:} All entropy densities are taken from ref.~\cite{Egan:2010}. The total}
       \hbox{entropies are based on the updated estimate of the proper volume}
       \hbox{of the observable Universe in Eq.~(\ref{eq:propvol}). The uncertain massive halo}
       \hbox{BHs \cite{Frampton:2009a} are not counted in the budget.}}}
 \end{tabular}\label{table:entropy}
}
\end{table*}

It is important for us to stress here that this is a new, key result. Ironically, many
other studies of horizon entropy (as we shall see shortly) have concluded that the
`correct' horizon to use is the apparent (or gravitational) horizon, not the particle
or event horizons, though with limited fundamental motivation. The consensus has
reached this point largely because the use of the other candidates simply does
not produce sensible (or even correct) first and second laws of horizon thermodynamics
\cite{Collins:1992,Hayward:1999,Hayward:1998,Bak:2000,Bousso:2005,Nielsen:2009}.
That the thermodynamics associated with the event horizon is ill-defined was further
argued in refs.~\cite{Davies:1988,Frolov:2003,Wang:2006}. But here, for the first time,
we are presenting a theoretical argument supporting the use of $R_{\rm  h}$ for such
purposes. It simply has to do with the fact that the proper size of the observable
Universe is $R_{\gamma,\,{\rm max}}$, which is indeed proportional to $R_{\rm h}$.
More strictly, one should use $\eta R_{\rm h}$ when evaluating the horizon entropy
and temperature rather than $R_{\rm h}$, as we shall discuss in \S~3.B, but the
proportionality constant $\eta$ has no impact on the first and second laws.

The use of the particle horizon, $R_{\rm p}$ (Eq.~\ref{eq:horpart}), to represent
the proper size of the observable Universe, instead of $R_{\gamma,\,{\rm max}}$, introduces
a non-negligible error in the estimated entropy budget (Table~\ref{table:entropy}).
According to ref.~\cite{Egan:2010}, the particle horizon today is $R_{\rm p}(t_0)=46.9\pm0.4$
Glyr, assuming a Hubble constant $H_0=70.5\pm1.3$ km s$^{-1}$ Mpc$^{-1}$. By comparison,
$R_{\rm h}(t_0)/2\approx 6.9$ Glyr for the same cosmological parameters. The observable
spherical volume estimated in previous works was therefore too big by a factor $\sim 300$.
Here, we adopt the latest {\it Planck} optimized parameters \cite{Planck:2018}, and estimate
\begin{equation}
R_{\gamma,\,{\rm max}}\approx (5.8\pm0.04)\left({\eta\over 0.4}\right)\;{\rm Glyr}\;,
\end{equation}
for a Hubble constant $H_0=67.4\pm0.5$ km s$^{-1}$ Mpc $^{-1}$. The value $\eta=0.4$ represents
the maximum proper distance traveled by photons reaching us in a $\Lambda$CDM background
cosmology \cite{Bikwa:2012}. The corresponding proper spherical volume of the observable
Universe, in the context of $\Lambda$CDM, is therefore
\begin{eqnarray}
V_{\rm obs}&=& 817\pm17\;{\rm Glyr}^3\cr
&=&(6.9\pm0.1)\times 10^{77}\;{\rm m}^3\;.\label{eq:propvol}
\end{eqnarray}

The cosmic entropy budget has contributions from many sources, most prominently from
black holes, the CMB and the relic neutrino background. Though the various workers
\cite{Kolb:1981,Frautschi:1982,Penrose:2004,Bousso:2007,Frampton:2008,Frampton:2009,Egan:2010}
who have attempted to estimate these quantities have not always agreed on their absolute
magnitude, their importance on a relative scale has never been in doubt. Adopting the estimates
from one of the most recent, and presumably most accurate, compilations, we show the dominant
contributions in Table~\ref{table:entropy}, together with their tentative sum. In this collection,
the entropy densities (analogous to Eq.~\ref{eq:CMBentropydensity} for the CMB) are taken
directly from ref.~\cite{Egan:2010}, but the total entropies have been updated using
our more physically motivated proper volume (Eq.~\ref{eq:propvol}) of the observable
Universe.

According to Table~\ref{table:entropy}, the largest contributions to the entropy of the
observable Universe today are made by supermassive black holes (see \S~3.B below). The
entropy of the CMB, a blackbody distribution with a present temperature $T_\gamma\sim 2.7$~K,
is at least $\sim 10^{14}$ times smaller. The entropy of any non-thermal radiation, including
light produced by stars and the interstellar medium, is even smaller---by a factor $\sim 10^3$
\cite{Frautschi:1982,Bousso:2007,Frampton:2009}.

But note that according to the discussion following Equation~(\ref{eq:CMBentropydensity}),
$S_{\rm CMB}$ has remained constant since decoupling (at $z_{\rm dec}\sim 1080$), corresponding
to cosmic time $\sim 380,000$ years. As we understand it, there were no black holes present
prior to the onset of re-ionization at $z\sim 15$, so according to this simple-minded
scenario, the total entropy ($S_{\rm bulk}$) in the bulk must have experienced several phases
of rapid, and dramatic, growth: first in reaching $S_{\rm CMB}$ at $\sim 380,000$ years,
then increasing to $S_{\rm bulk}$ today, starting near the onset of Pop III star formation
at $t\sim 200-400$ Myr.

How much confidence should we place in this conclusion, however?  Gravitational entropy probably
includes a contribution from the gravitational field itself, but its nature and size have remained
a mystery. Penrose \cite{Penrose:1987} proposed that it may be given by the Weyl curvature tensor,
$W_{\alpha\beta\gamma\delta}$, which provides a measure of the curvature of spacetime (and hence
the `strength' of the gravitational field). In general relativity, $W_{\alpha\beta\gamma\delta}$
is the only part of the curvature that exists in free space, corresponding to solutions of the
vacuum Einstein equations. Unlike the Ricci curvature $R_{\mu\nu}$, which vanishes in the absence
of matter, the Weyl curvature may still be non-zero in vacuum if, e.g., gravitational waves are
propagating through the medium. It therefore seems to be a more appropriate choice for the
representation of gravitational entropy, which may be non-zero even in `empty' space, but
Table~\ref{table:entropy} does not include such a contribution to the entropy budget because
it is difficult to quantify.

Once matter fluctuations began to grow, and the spacetime became clumpy,
$W_{\alpha\beta\gamma\delta}$ grew as well, as did the gravitational entropy.
This process culminated with the large nonlinear overdensities we see today. In extreme
cases, the clumping led to the formation of black holes, whose entropy is well known
(Table~\ref{table:entropy}). The story with gravitational entropy is therefore incomplete.
It appears that we know the end point fairly well, but we have trouble quantifying the
overall contribution of gravity to the total entropy budget prior to the formation of
supermassive black holes. Was gravitational entropy in the early Universe much smaller
than it is today, or was it merely present in a different form, i.e., gravitational
fields as opposed to fully formed black holes? It is hard to say, but our analysis in
\S~4 below may provide some helpful clues.

\subsection{Entropy of the Apparent Horizon}
The general discussion concerning cosmic entropy includes a growing realization
(some would say `acceptance') that a complete understanding of FLRW thermodynamics
ought to include both the entropy in the bulk, as summarized in \S~3.A above, and
the entropy associated with the apparent (or gravitational) horizon, $R_{\rm h}$.
This horizon has both a temperature, $T_{\rm h}$, and an entropy, $S_{\rm h}$,
extending the basic properties of black-hole horizons discovered in the 1970's
\cite{Bekenstein:1973,Hawking:1970,Hawking:1975}. The temperature is inferred
using the Hamilton-Jacobi variant of the Parikh-Wilczek `tunneling' approach
\cite{Parikh:2000}, though different definitions of `surface' gravity produce
some variation on the actual value of $T_{\rm h}$. The Kodama-Hayward version
is noteworthy because it is based on a conserved current even when a timelike
Killing vector is absent \cite{Kodama:1980}. In addition, the Noether charge
associated with the Kodama vector is the Misner-Sharp-Hernandez mass
\cite{Misner:1964,Hernandez:1966}, which is commonly used to represent
the total internal energy contained within a sphere of radius $R_{\rm h}$
\cite{Melia:2007}.

Coupling the entropy in the bulk to the entropy of a horizon actually has
a historical precedent dating back to Einstein himself \cite{Einstein:1997}.
It is well known that the Einstein-Hilbert action can be decomposed into
a bulk term and surface term (see also ref.~\cite{Padmanabhan:2010}). Yet
the field equations (and their solutions) may be derived exclusively from
the variations of the bulk, without any recourse to the surface term. The
reason for this, it turns out, is that the bulk and surface terms are actually
not independent of each other \cite{Padmanabhan:2002}. The term `holographic'
(see \S~2.B above) was thus coined because the information about the bulk
action functional is encoded in the boundary action functional.

The first exploration of FLRW horizon thermodynamics was made in de Sitter space,
first by Gibbons and Hawking \cite{Gibbons:1977}, and later by
refs.~\cite{Birrell:1982,Davies:1986,Medved:2002,Parikh:2002}, though in terms of
the cosmic event horizon which, in de Sitter space, is static (see Eq.~\ref{eq:horev}).
Its temperature and entropy (commonly referred to as {\it Gibbons-Hawking} entropy) are,
respectively,
\begin{equation}
T_{\rm e}\equiv {1\over k_{\rm B}}{\hbar H\over 2\pi}\;,\label{eq:deStemp}
\end{equation}
and
\begin{equation}
S_{\rm e}\equiv {k_{\rm B}c^3\over\hbar G}{A_{\rm e}\over 4}=
{k_{\rm B} c^5\over \hbar G}{\pi\over H^2}\;,\label{eq:GHentropy}
\end{equation}
analogous to those of a Schwarzschild event horizon, originally inferred using
Euclidean field theory techniques \cite{Gibbons:1977}. In these expressions,
$k_{\rm B}$ is the Boltzmann constant, $H$ is the Hubble parameter (which is constant
in de Sitter space), $G$ is the gravitational constant, and
\begin{equation}
A_{\rm e}\equiv 4\pi R_{\rm e}^2
\end{equation}
is the area of the event horizon written in terms of its radius $R_{\rm e}$.
Note, however, that in de Sitter space the event and apparent horizons
are degenerate since $H$ (and therefore $R_{\rm h}$) is constant. So for de Sitter,
$R_{\rm e}=R_{\rm h}\equiv c/H$. As we shall see, the horizon area relevant to all
FLRW cosmologies can therefore be written more commonly as
\begin{equation}
A_{\rm h}\equiv 4\pi R_{\rm h}^2\;.\label{eq:Ah}
\end{equation}

A more physically motivated way of writing the entropy, highlighting the notion that it
represents the number of units of ``quantum area" that fit within $A_{\rm h}$, is
\begin{equation}
S_{\rm h}=k_{\rm B}{A_{\rm  h}\over 4 \ell_{\rm Pl}^2}\;,\label{eq:appentropy}
\end{equation}
where
\begin{equation}
\ell_{\rm Pl}\equiv\sqrt{G\hbar\over c^3}
\end{equation}
is the Planck length. If the cosmic spacetime today were de Sitter, the (event) horizon would
produce thermal radiation with a characteristic temperature $T_{\rm  e}\sim 3\times 10^{-30}$ K,
and its entropy would be constant, given that the de Sitter spacetime curvature is independent
of time.

A broader discussion relevant to the thermodynamical properties of the apparent horizon,
the Kodama vector and other issues associated with FLRW in general, may be found in
ref.~\cite{DiCriscienzo:2010}. The Kodama-Hayward temperature of the apparent horizon is given by
\begin{equation}
T_{\rm h}={1\over k_{\rm B}}{\hbar c\over 8\pi R_{\rm h}}(1-3w)\;,\label{eq:KHtemp}
\end{equation}
where $w\equiv p/\rho$ is the equation-of-state of the cosmic fluid, in terms
of its total energy density $\rho$ and total pressure $p$. One can easily confirm that
Equation~(\ref{eq:KHtemp}) reduces to Equation~(\ref{eq:deStemp}) in the limit $w\rightarrow -1$.
Correspondingly, it is not difficult to understand from Equation~(\ref{eq:appentropy}) that the
apparent horizon entropy in an expanding Universe increases if $\rho+p>0$ (which produces a
monotonically decreasing Hubble parameter $H[t]$), remains constant when $p=-\rho$ (i.e., de
Sitter space, as we have seen), and decreases if $p<-\rho$, which produces
a monotonically decreasing Hubble radius $R_{\rm h}$ and apparent horizon surface area
$A_{\rm h}$ (see Eq.~\ref{eq:Ah}).

\section{First and Second Laws of Thermodynamics}
Let us now see how the first and second laws of thermodynamics must be framed in FLRW
in terms of the various physical concepts introduced in \S~3 above, and why cosmic
entropy is simplified in the $R_{\rm h}=ct$ universe compared to $\Lambda$CDM.

The total entropy within the visible Universe, whose proper size is
$R_{\gamma,\,{\rm max}}=\eta R_{\rm h}$, may be written
\begin{equation}
S=S_{\rm bulk}+S_{\rm h}\;,\label{eq:totalS}
\end{equation}
where $S_{\rm bulk}$ includes all of the elements in Table~\ref{table:entropy}, and
$S_{h}$ is given in Equation~(\ref{eq:appentropy}). As noted earlier, $\eta=1$ for de
Sitter space, but $\eta\lesssim 0.5$ for all other FLRW cosmologies expanding from an
initial singularity at the Big Bang. In all cases, however, it is the physics associated
with the gravitational (or {\it apparent}) horizon that determines the properties of
$S_{\rm bulk}$ and $S_{h}$, not the particle or event horizons. As we shall see shortly,
the distinction between $R_{\gamma,\,{\rm max}}$ and $R_{\rm h}$ in calculating $S_{\rm bulk}$
does not affect our analysis. The primary purpose of introducing the former radius (via
the Appendix) as the proper size of the visible Universe is to provide greater
justification for the relevance of $R_{\rm h}$ to the question of cosmic entropy.

In spherically-symmetric metrics such as FLRW, the properties of the
apparent horizon may be determined using the previously introduced Misner-Sharp-Hernandez mass
\cite{Misner:1964,Hernandez:1966,Melia:2007,Melia:2018b}, which coincides with the Hawking-Hayward
quasi-local energy in such systems \cite{Hawking:1968,Hayward:1994}. Under such circumstances, it
is given simply as
\begin{equation}
M_{\rm h}\equiv {4\pi\over 3c^2}{R_{\rm h}}^3\rho\;.\label{eq:Mh}
\end{equation}
It is not difficult to show \cite{Melia:2007,Melia:2009} that a second useful relation is therefore
\begin{equation}
R_{\rm h}={2GM_{\rm h}\over c^2}\;,\label{eq:Rh}
\end{equation}
and this in turn gives $R_{\rm h}=c/H$ in a spatially flat Universe. As we shall confirm shortly,
one thus finds that the first and second laws of thermodynamics applied to de Sitter space are
satisfied automatically, since the internal energy $M_{\rm h}c^2$, the volume $V_{\rm h}\equiv 
4\pi R_{\rm h}/3$, and the horizon entropy $S_{\rm h}$, are all constant, so $dS_{\rm h}$,
$dM_{\rm h}c^2$ and $dV_{\rm h}$ vanish uniformly.

For the broader class of FLRW cosmologies, we follow
refs.~\cite{Hayward:1998,Hayward:1999,Bak:2000} and define the so-called {\it work density}
\begin{equation}
W\equiv -{1\over 2}g_{ab}T^{ab}\;,\label{eq:workdensity}
\end{equation}
and the energy flux across the apparent horizon
\begin{equation}
\Psi_a\equiv {T_a}^b\partial_b [a(t)r]+W\partial_a[ar]\;,\label{eq:energysupply}
\end{equation}
where $g_{ab}$ are the FLRW metric coefficients in Equation~(\ref{eq:FLRW}), $T^{ab}$
is the energy-momentum tensor for a perfect fluid, $a(t)$ is the expansion factor and
$r$ is the comoving radius, and the indices `a' and `b' run over the values $(0,1)$.
The work density represents the work done by a change in the horizon's radius, while
the energy-supply vector, $A_{\rm h}\Psi_a$, represents the total energy flow through
that horizon. For a perfect fluid, it is not too difficult to show that
\begin{equation}
W={\rho-p\over 2}
\end{equation}
where, as usual, $\rho$ and $p$ are the total energy density and pressure.

The Einstein equations then lead to the expression
\begin{equation}
\nabla_a\, M_{\rm h}c^2=A_{\rm h}\Psi_a+W\,\nabla_a V_{\rm h}\;,\label{eq:UFL}
\end{equation}
known as the `unified first law' \cite{Hayward:1998,Hayward:1999}. The interpretation
of this equation is that the energy supply, $-A_{\rm h}\Psi_a$, across the apparent
horizon is the change in heat,
\begin{equation}
\nabla_a\,Q_{\rm h}=-A_{\rm h}\Psi_a=T_{\rm h}\,\nabla_a\,S_{\rm  h}\;,\label{eq:heat}
\end{equation}
and that this heat goes into changing the internal energy
\begin{equation}
E_{\rm h}\equiv M_{\rm h}c^2\;,\label{eq:enbulk}
\end{equation}
and performing work due to the change in size of this horizon. In writing Equation~(\ref{eq:heat}),
we have used the idea that the Universe and its horizon entropy encode the
positive heat out thermodynamic sign convention (see, e.g., ref.~\cite{Tian:2015}).
The fact that the coefficient multiplying $\nabla_a V_{\rm h}$
in the work term of Equation~(\ref{eq:UFL}) is not simply $p$ is apparently due to the
fact that $R_{\gamma,\,{\rm max}}$ is not a comoving radius \cite{Faraoni:2011}.

As a basic test of the validity of these various ideas, let us follow the rather
straightforward demonstration that Equation~(\ref{eq:heat}) yields an evolution in
the horizon entropy fully consistent with its definition in Equation~(\ref{eq:appentropy}).
From Equations~(\ref{eq:Rh}) and (\ref{eq:enbulk}), we see that a change in the
horizon radius by an amount $dR_{\rm h}$ produces a change to the internal energy of
\begin{equation}
dE_{\rm h}={c^4\over 2G}\,dR_{\rm h}\;.
\end{equation}
Correspondingly, the work done during this change is
\begin{equation}
W\,dV_{\rm h}=(1-w){3c^4\over 4G}\,dR_{\rm h}
\end{equation}
where, as always, $w\equiv p/\rho$ and we have assumed a spatially flat FLRW spacetime.
Equations~(\ref{eq:UFL}) and (\ref{eq:heat}) therefore suggest that
\begin{equation}
T_{\rm h}\,dS_{\rm h}={c^4\over 2G}\left({1-3w\over 2}\right)\,dR_{\rm h}\;,\label{eq:TdS}
\end{equation}
and one can trivially confirm that this expression is completely consistent with
Equations~(\ref{eq:appentropy}) and (\ref{eq:KHtemp}), {\it regardless of the value
of} $w$. The extension of Bekenstein and Hawking's work on black-hole horizon
entropy \cite{Bekenstein:1973,Hawking:1970,Hawking:1975} to the cosmological
context therefore appears to be fully intact and completely consistent with
Einstein's equations.

The corresponding thermodynamic quantities pertaining to the bulk satisfy the traditional
Gibbs equation
\cite{Izquierdo:2006,Dutta:2010}
\begin{equation}
T_{\rm bulk}\,dS_{\rm bulk}=dE(R_{\gamma,\,{\rm max}})+p\,dV(R_{\gamma,\,{\rm max}})\;.\label{eq:Gibbs}
\end{equation}
But whereas Equation~(\ref{eq:TdS}) is satisfied by all cosmologies, we shall see shortly
that the evolution of the physical entropy $S_{\rm bulk}$ in $\Lambda$CDM, based solely on
Equation~(\ref{eq:Gibbs}), departs dramatically from the desired nondecreasing behavior,
completely at odds with our discussion in \S\S~1--3 above. The growing consensus 
today is that one must therefore resort to a {\it generalized} second law (GSL),
in which the geometrically defined $S_{\rm h}$ must be added to $S_{\rm bulk}$ in order
to create a nondecreasing {\it total} entropy $S$, as given in Equation~(\ref{eq:totalS}). This
approach is not universally accepted, however, due to its reliance on the `local equilibrium
assumption' (see below) which posits that the interior region and the apparent horizon have the same
temperature, $T_{\rm bulk}=T_{\rm h}$ \cite{Setare:2007,Karami:2010,Abdolmaleki:2014,Herrera:2014}.
The principal goal of this paper is to demonstrate that this difficulty (perhaps one should
say, `unmotivated requirement') is completely absent in the $R_{\rm h}=ct$ universe.

\subsection{First and Second Laws in $\Lambda$CDM}
From the Gibbs Equation~(\ref{eq:Gibbs}) and the definition of $T_{\rm bulk}$ and
$S_{\rm bulk}$, one can see that
\begin{equation}
T_{\rm bulk}\,dS_{\rm bulk}={c^4\over 2G\eta^3}\,dR_{\rm h}\left(1+3w\right)\;,\label{eq:Sbulk}
\end{equation}
which becomes our first law for the bulk. The hurdle faced by $\Lambda$CDM (see \S\S~2 and 3)
in accounting for the cosmic entropy evolution with time is therefore clearly apparent, since
any acceleration of the universal expansion requires $\rho+3p<0$, i.e., $w<-1/3$. Clearly,
\begin{equation}
{dS_{\rm bulk}\over dR_{\rm h}}<0\quad (\forall\;\, w<-1/3)
\end{equation}
for such a model, regardless of how one interprets $T_{\rm bulk}$. The second law is therefore
not satisfied for the bulk in the context of $\Lambda$CDM.

The GSL was introduced specifically to overcome this deficiency (\S~3.B), with the
hope that $dS/dR_{\rm h}\ge 0$ (see Eq.~\ref{eq:totalS}), in spite of the fact that
the bulk cosmic entropy in the standard model is apparently decreasing with time when
the expansion is accelerated. The inspiration for this idea came by way of the GSL in
black-hole thermodynamics \cite{Bekenstein:1973}, where the event horizon entropy added
to the entropy outside the black hole never decreases.

Let use see what is required for the GSL to work in this fashion. It is straightforward
to show from Equations~(\ref{eq:TdS}) and (\ref{eq:Gibbs}) that the combined cosmic entropy
in $\Lambda$CDM never decreases as long as
\begin{eqnarray}
dS_{\rm h}+dS_{\rm bulk}&=&{c^4\over 2G}\,dR_{\rm h}\left({1-3w\over 2T_{\rm h}}
+{1+3w\over \eta^3 T_{\rm bulk}}\right)\nonumber\\
&\ge& 0\;.\label{eq:GSL}
\end{eqnarray}
That is, the GSL is satisfied as long as
\begin{equation}
w\le {1\over 3}{1+\alpha\over 1-\alpha}\;,\label{eq:alphaconst}
\end{equation}
where
\begin{equation}
\alpha\equiv {2\over \eta^3}\left({T_{\rm h}\over T_{\rm bulk}}\right)\;.\label{eq:alpha}
\end{equation}
Evidently, the GSL is satisfied for any FLRW cosmology with an equation-of-state $w\le 1/3$
in the cosmic fluid, as long as $\alpha <1$. Otherwise,
the Universe must be accelerating and $\alpha\gg 2$ to comply with the requirement
that $dS\ge 0$ at all times. It is this tight coupling between $T_{\rm bulk}$ and $T_{\rm h}$
that has promoted the so-called `local equilibrium assumption' introduced in the
discussion following Equation~(\ref{eq:Gibbs}) above. More precisely, the constraint in
Equation~(\ref{eq:alphaconst}) seems to require a proportionality between $T_{\rm h}$ and
$T_{\rm bulk}$, though the proportionality constant is often assumed to be unity
\cite{Setare:2007,Karami:2010,Abdolmaleki:2014,Herrera:2014}. It needs to be
acknowledged, however, that there is no known fundamental mechanism responsible
for maintaining the bulk in thermal equilibrium with the horizon. Indeed, this condition
cannot be satisfied at all if $T_{\rm bulk}$ is associated primarily with the CMB
blackbody temperature $T_{\rm BB}$, as we shall see below.

\subsection{First and Second Laws in $R_{\rm h}=ct$}
The $R_{\rm h}=ct$ cosmology has been under development for over 15 years now, largely
motivated by observational constraints, but more recently finding significant fundamental
support from renewed interest in the FLRW metric itself \cite{Melia:2019a}. A complete
accounting of this model may be found in a recently released monograph \cite{Melia:2020a}.
A quick summary of its observational tests may also be found in Table~2 of
ref.~\cite{Melia:2018a}. Like $\Lambda$CDM, $R_{\rm h}=ct$ is a FLRW cosmology with
a cosmic fluid comprised of matter (baryonic and dark), radiation and dark energy, though
the latter is not a cosmological constant. The key new physical input that distinguishes
it from the standard model is the so-called zero active mass condition in general relativity,
\begin{equation}
\rho+3p=0\;,\label{eq:EoS}
\end{equation}
sustained throughout the expansion. In other words, $R_{\rm h}=ct$ is essentially $\Lambda$CDM,
though constrained by the equation-of-state $w=-1/3$ over the entire cosmic evolution.

To understand why it is now timely to examine the issue of cosmic entropy in this
cosmology, let us list some of the notable features that have allowed it to overcome
many major hurdles and inconsistencies in the standard model. This is only a small
sample; its successes extend well beyond this brief summary. First and foremost,
$R_{\rm h}=ct$ has no horizon problems. Whereas models with an early phase of deceleration,
such as $\Lambda$CDM, must find ways of explaining how regions of the sky beyond their causal
horizon achieved similar physical conditions, this issue is completely absent in $R_{\rm h}=ct$
\cite{Melia:2013b}, which has always expanded at a constant rate. Inflation was introduced in
part to address the CMB temperature horizon problem in the standard model
\cite{Starobinsky:1980,Guth:1981,Linde:1982}, but has yet to find a completely self-consistent
theoretical framework. The electroweak horizon problem appears to be worse. As of today, no viable
explanation has been found for how opposite sides of the Universe seemingly acquired the same
Higgs vacuum expectation value. But again, this problem is completely absent in $R_{\rm h}=ct$
\cite{Melia:2018c}.

The $R_{\rm h}=ct$ cosmology apparently also avoids the so-called trans-Planckian
problem in $\Lambda$CDM
\cite{Martin:2001,Brandenberger:2001,Brandenberger:2013,Melia:2019b,Melia:2020b}.
Whereas quantum fluctuations in the standard model would have been seeded in the
Bunch-Davies vacuum, well below the Planck scale, they emerged into the semi-classical
Universe right at the Planck scale, at about the Planck time, in $R_{\rm h}=ct$. Our
current physical theories are not valid in the Planck regime, so it is currently a mystery
exactly how these fluctuations evolved into the semi-classical Universe in the context
of standard inflationary cosmology.

The third and final example we mention here has to do with the timeline in $\Lambda$CDM,
which appears to be overly compressed at large $z$, based on the observation of high-redshift
quasars and the so-called `too-early' appearance of galaxies \cite{Steinhardt:2016}
and the accelerated rate of structure formation in the early Universe. In contrast, the
time-redshift relation in $R_{\rm h}=ct$ matches the rate of supermassive black-hole growth
very well \cite{Melia:2013c,Melia:2018d}, and readily accounts for the appearance of galaxies
\cite{Melia:2014} and large halos \cite{Yennapureddy:2019} at $z\gtrsim 10$.

One of the few remaining issues to consider is the nature and evolution of cosmic entropy in
the $R_{\rm h}=ct$ cosmology, which we now address. We see right away from Equations~(\ref{eq:Sbulk})
and (\ref{eq:EoS}) why cosmic entropy is greatly simplified in this model compared to $\Lambda$CDM.
Evidently,
\begin{equation}
S_{\rm bulk}={\rm constant}\quad (w=-1/3) 
\end{equation}
throughout the cosmic expansion, independently of how we choose to evaluate $T_{\rm bulk}$.
There is no need to `fix' an entropy problem by introducing a GSL, except to determine the
arrow of time \cite{Carroll:2004} which, in this model, is simply identified from the expansion
or contraction of the apparent horizon. Insofar as $S_{\rm bulk}$ is concerned, however, there
is no need to explain an anomalously low entropy at the beginning---because it wasn't low---nor
is there any concern that it may actually be decreasing with time---since it is always constant---which
would otherwise violate the second law. In this picture, the various contributions to the total entropy
in the bulk shift relative to each other as the Universe expands, but always in such a way as to maintain
a constant global value established at the beginning.

The GSL is not required in this model, but if we were to introduce it analogously to the procedure we
followed in arriving at Equation~(\ref{eq:GSL}) for $\Lambda$CDM, then the total cosmic entropy
(Eq.~\ref{eq:totalS}) would grow as
\begin{equation}
S=k_{\rm B}{\pi c^2\over \ell^2_{\rm Pl}}t^2\;,
\end{equation}
affirming the idea floated earlier that the arrow of time in this cosmology might be determined by
the expansion of its apparent horizon.

In contrast to the standard model, one can see from this result that it is not necessary to
speculatively constrain the evolution of $T_{\rm bulk}$ in $R_{\rm h}=ct$. Ironically, however, it
would actually be easier to demonstrate that $T_{\rm bulk}\propto T_{\rm h}$ in $R_{\rm h}=ct$ than
in $\Lambda$CDM, if one interprets $T_{\rm bulk}$ as being primarily $T_{\rm BB}$. This is evident
from Equation~(\ref{eq:KHtemp}) and the CMB temperature $T_{\rm BB}$ which scale, respectively, as
\begin{equation}
T_{\rm h}(z)={1\over k_{\rm B}}{\hbar c\over 4\pi\,R_{\rm h}}
={1\over k_{\rm B}}{\hbar c\over 4\pi\,R_{\rm h}(0)}(1+z)\;,
\end{equation}
and
\begin{equation}
T_{\rm BB}(z)=T_0(1+z)\;,
\end{equation}
where $R_{\rm h}(0)$ is the radius of the apparent horizon today and $T_0$ is the current
CMB temperature. That is,
\begin{equation}
T_{\rm bulk}=b\,T_{\rm h}\;,\label{eq:Tprop}
\end{equation}
where
\begin{equation}
b\equiv k_{\rm B}{4\pi R_{\rm h}(0)T_0\over\hbar c}\approx 2.1\times 10^{30}\;.
\end{equation}
We stress, however, that though these temperatures scale in proportion with the expansion of the
Universe, Equation~(\ref{eq:Tprop}) does not at all require that the bulk be in equilibrium with
the apparent horizon, as is often assumed in the context of $\Lambda$CDM.

\section{Impact on Theories of the Early Universe}
Like $\Lambda$CDM, the $R_{\rm h}=ct$ universe contains baryonic and dark matter ($\rho_{\rm m}$),
radiation ($\rho_{\rm r}$) and dark energy ($\rho_{\rm de}$). Unlike the standard model, however,
these densities must exist in proportions consistent with the zero-active mass condition,
$\rho+3p=0$, where $\rho=\rho_{\rm m}+\rho_{\rm r}+\rho_{\rm de}$ and 
$p=p_{\rm r}+p_{\rm m}+p_{\rm de}$. With $p_{\rm m}=0$, these imply that
$p_{\rm r}+p_{\rm de}=-(\rho_{\rm m}+\rho_{\rm r}+\rho_{\rm de})/3$. It is not difficult to
convince oneself that this equation-of-state can be satisfied if $w_{\rm de}\equiv 
p_{\rm de}/\rho_{\rm de}=-1/2$, with a fractional representation of $\rho_{\rm de}/\rho=1/3$
and $\rho_{\rm m}/\rho=2/3$ in the local universe ($z\sim 0$) and 
$\rho_{\rm de}/\rho=0.8$ and $\rho_{\rm r}/\rho=0.2$ as $z\rightarrow\infty$ (see, e.g.,
ref.~\cite{Melia:2016,Melia:2020a}. Quite clearly, dark energy could not be a cosmological
constant in such a model, and one might go further and speculate that a fraction of dark
energy present at the beginning eventually `decayed' to create matter with a fractional
representation of $1/3$ towards the present day. Evidently, both dark matter and dark energy
would thus be considered extensions to the standard model of particle physics. 

We don't know yet, though, whether radiation and dark energy were present from the very
beginning, or whether they were preceded by something else, perhaps a scalar field. But
in order for this field to also satisfy the zero-active mass condition, it would have to 
have had an exponential potential \cite{Melia:2019b} reminiscent of the category of minimally 
coupled fields explored in the 1980’s, designed to produce so-called power-law inflation 
\cite{Abbott:1984,Lucchin:1985,Barrow:1987,Liddle:1989}. Unlike the other members of
this class, however, this zero-active mass field actually would not have inflated,
since $a(t)=(t/t_0)$. 

But given these attributes, a more `natural' candidate for the incipient dominant content of
the Universe could very well have been cosmic strings \cite{Dabrowski:1989} arising from
phase transitions in such a field. The tension along a string is equal to its energy per 
unit length \cite{Zeldovich:1980,Vilenkin:1981}, so the equation-of-state of a chaotic 
ensemble of strings is simply $p=-\rho/3$, as required in $R_{\rm h}=ct$. Though such 
strings are unlikely to be dynamically relevant today, they may have been dominant at 
early times \cite{Kibble:1976}. The earliest manifestation of the $R_{\rm h}=ct$ cosmology 
could thus have been a string-dominated universe, that eventually evolved into the more 
`standard' scenario we see today.

The possible consequences of such a beginning are numerous and intriguing. Certainly, if
$R_{\rm h}=ct$ correctly describes the background spacetime of the Universe, the total
entropy in the bulk has remained constant since the beginning, implying that a reasonable
`measurement' of $S_{\rm bulk}$ today would directly reveal an important attribute of
the physical system during the string-dominated era---without us having to worry about
the details concerning the transition from those earliest times to the subsequent 
more `standard' evolution with $\rho=\rho_{\rm m}+\rho_{\rm r}+\rho_{\rm de}$. For example,
such a constraint would directly reveal whether entropy variations in the equation-of-state of 
the cosmic strings in the cosmological fluid, due to variations in the local density and
gravitational radiation, could have led to perturbations of the background matter density, 
and the subsequent cosmic-string induced formation of structure \cite{Avelino:1996}.
Such ideas and possibilities will be explored elsewhere.

\section{Conclusion}
Prior to the introduction of the GSL, there was considerable confusion about how to handle
cosmic entropy in the context of $\Lambda$CDM. The notion that $S_{\rm bulk}$ should have
an anomalously low initial value was considered to be so `unnatural' that some rather exotic
(and equally `unnatural') solutions were proposed. The idea that the Universe ought to have
such a condition both at the beginning and at the `end' was considered a viable possibility
\cite{Gold:1992,Price:1996}. But all such models appear to be very {\it ad hoc}, actually
exacerbating the `unnaturalness' of the boundary conditions. The GSL at least obviates the
need to patch the standard model in such a poorly motivated fashion, but it may suffer from
an inconsistency of its own when applied to $\Lambda$CDM.

Busso \cite{Busso:1999} conjectured that the horizon enclosing the bulk should be a lightlike
hypersurface, building on the argument of Fischler \& Susskind \cite{Fischler:1998}, who used
lightlike hypersurfaces to relate entropy and area (Eq.~\ref{eq:appentropy}). Since then,
however, several workers have established that neither the particle
horizon (Eq.~\ref{eq:horpart}), nor the event horizon (Eq.~\ref{eq:horev}), satisfy the laws of
thermodynamics. Attention has thus been redirected onto the apparent (or gravitational) horizon
to fulfill this role, and we describe in the Appendix a possible fundamental reason why it
has to be this way. But the apparent horizon is generally not null---except in one unique
cosmology: the $R_{\rm h}=ct$ universe. This model is characterized by the zero active mass
equation-of-state (Eq.~\ref{eq:EoS}), for which the radius $R_{\rm h}$ increases at lightspeed,
i.e., $dR_{\rm h}/dt=c$---hence the eponymous naming of the model. Thus, the $R_{\rm h}=ct$
cosmology uniquely satisfies the second law of thermodynamics in the bulk, and horizon
thermodynamics on the null boundary hypersurface (at $R_{\rm h}$) consistent with the
proper size of the observable Universe.

The fact that $S_{\rm bulk}$ is constant in the $R_{\rm h}=ct$ universe actually solves
two problems in standard cosmology: (1) it eliminates the need for the Universe to have begun
its expansion with an anomalously low entropy, and (2) it explains how the cosmic entropy
could have been so large by the time of decoupling to produce the CMB. A fixed value for
$S_{\rm bulk}$ is also more in line with the cosmological principle, which posits that
the Universe is isotropic and homogeneous on large scales (certainly larger than $\sim 300$ Mpc).
One should not expect to see a net inflow (outflow) of entropy into (out of) a finite volume
if the physical conditions are the same everywhere on each given time slice.

We thus see that the $R_{\rm h}=ct$ cosmology may have resolved yet another difficulty or
inconsistency in $\Lambda$CDM. Like several of the other problems preceding it, the IEP
has been with us for several decades, stubbornly resisting attempts at reconciling the
hypothesized expansion history in the standard model with expectations based on
established physical theories. If the solution presented in this paper turns out to be
correct, the answer is actually elegant and straightforward, requiring no `unnatural'
initial conditions, nor any forced thermodynamic equilibrium between the bulk and its
apparent horizon. One might say the answer seems quite natural after all.

\section*{APPENDIX \\ Proper Size of the Observable Universe}
Standard cosmology is based on the Friedmann-Lema\^{i}tre-Robertson-Walker (FLRW) metric for a
spatially homogeneous and isotropic three-dimensional space, expanding or contracting
as a function of time:
\begin{equation}
ds^2=c^2\,dt^2-a^2(t)\left[{dr^2\over (1-kr^2)}+r^2d\Omega^2\right]\,,\label{eq:FLRW}
\end{equation}
\vskip 0.1in\noindent
where $d\Omega^2\equiv d\theta^2+\sin^2\theta\,d\phi^2$. In these coordinates, $t$ is the
cosmic time, measured by a comoving observer (and is independent of position), $a(t)$ is
the universal expansion factor, and $r$ is the comoving radius. The geometric factor $k$ is
$+1$ for a closed universe, $0$ for a flat universe, and $-1$ for an open universe. Most
of the data suggest that the Universe is spatially flat, so we assume $k=0$ throughout this paper.

It is not difficult to show \cite{Melia:2013a,Melia:2018b} from Equation~(\ref{eq:FLRW}) and the
definition of proper distance, i.e., $R\equiv a(t)r$, that the null geodesic equation for a ray of
light propagating radially may be written
\begin{equation}
{dR_\gamma\over dt}=c\left({R_\gamma\over R_{\rm h }}\pm 1\right)\;,\label{eq:nullgeo}
\end{equation}
where $R_{\rm h}\equiv c/H$ is the radius of the gravitational (or apparent) horizon, and the sign
$\pm$ refers to either inward ($-$ sign) or outward ($+$ sign) propagation. The former corresponds
to null geodesics actually reaching the observer at the origin of coordinates, and are therefore
relevant in determining the proper size of the observable Universe \cite{Melia:2013a}. As we shall
see below, the outwardly propagating rays (with the $+$ sign in Eq.~\ref{eq:nullgeo}) define a
`particle' horizon which, however, cannot be used as a measure of the size of the observable
Universe because these null geodesics are directed away from the observer and never return
to their location.

From Equation~(\ref{eq:nullgeo}) for an inwardly directed ray (i.e., with a negative sign on the
righthand side), it is clear that $dR_\gamma/dt=0$ when $R_\gamma=R_{\rm h}$. In other words,
the spatial velocity of light measured in terms of the proper distance per unit cosmic time
vanishes at the gravitational (or `apparent') horizon. When $R_\gamma>R_{\rm h}$, the photon's
proper distance from us actually increases, even though the photon's velocity vector points
towards the origin. The photon approaches the observer only when it is located within their
apparent horizon at $R_{\rm h}$. This separation at $R_{\rm h}$ of null geodesics approaching
the observer from those receding is well known in general relativity, but it is even simpler
to understand in the cosmic context due to the fact that FLRW is spherically symmetric.

But note that since $R_{\rm h}$ is a function of time, ${\dot{R}}_\gamma$ flips sign
depending on whether $R_\gamma$ overtakes $R_{\rm h}$, or vice versa. For example, a photon
emitted beyond $R_{\rm h}(t_e)$ at some time $t_e<t$ (where $t$ is the present), begins
its trajectory receding from the observer, yet stops at $R_\gamma=R_{\rm h}$, and reverses
direction when $R_{\rm h}$ overtakes $R_\gamma$. Consequently, the path of the null geodesic
$R_\gamma(t)$ depends on the cosmology, because the expansion history is solely responsible
for the evolution in $R_{\rm h}$. This dependence of $R_\gamma(t)$ on the evolution of
$R_\gamma/R_{\rm h}$ affirms the nature of gravitational (or apparent) horizons discussed
in the literature \cite{Melia:2007,Ben-Dov:2007,Faraoni:2011,Bengtsson:2011,Faraoni:2015}.

The proper size of our visible Universe hinges on the solution to Equation~(\ref{eq:nullgeo})
for the radius $R_{\rm h}(t)$ associated with the chosen cosmology, starting at the Big Bang
($t=0$) and ending at the observer's time. It is determined by the greatest proper distance
achieved by those null geodesics that actually reach the observer at the origin of the
coordinates. No matter what, however, Theorem~1 in ref.~\cite{Melia:2018b} ensures that
the proper size of the visible Universe at time $t$ is always subject to the constraint
\begin{equation}
R_{\gamma,\,{\rm max}}\le R_{\rm h}(t)\;,
\end{equation}
in any cosmology expanding monotonically with $\dot{H}\le 0$.

Thus, no matter how $R_{\rm h}$ evolves in time, none of the light detected by the observer
at time $t$ originated from beyond their gravitational (or apparent) horizon at that time.
Indeed, detailed solutions to Equation~(\ref{eq:nullgeo}) for typical cosmological models
show that expanding universes have a visibility limit restricted to about half of the
current gravitational radius $R_{\rm h}(t)$, or even somewhat less \cite{Melia:2013a}.
The only exception is de Sitter, for which $R_{\gamma,\,{\rm max}}=R_{\rm h}(t)$ (see
below). One may therefore summarize these results with the following statement: the proper
size of the observable Universe is
\begin{equation}
R_{\gamma,\,{\rm max}}=\eta R_{\rm h}(t)\;,\label{eq:maxrad}
\end{equation}
where $\eta$ ($\sim 1/2$) is a cosmology-dependent constant.

It is not difficult to understand the physical reason behind this outcome \cite{Melia:2013a}.
In all models other than de Sitter, there were no pre-existing detectable sources away from
the origin of the observer's coordinates prior to the Big Bang. Thus, light reaching us at
time $t$ from the most distant sources was emitted only after the latter had reached their
farthest detectable proper distance, i.e., $\sim$~$R_{\rm h}(t)/2$.

Equation~(\ref{eq:maxrad}) needs to be contrasted with typical past treatments of the
cosmic entropy budget (see, e.g.,
refs.~\cite{Kolb:1981,Frautschi:1982,Penrose:2004,Frampton:2008,Frampton:2009,Egan:2010}),
in which the proper radius of the observable Universe was instead assumed to be the particle horizon,
\begin{equation}
R_{\rm p}(t)\equiv a(t)\int_0^t {c\,dt^\prime\over a(t^\prime)}\;.\label{eq:horpart}
\end{equation}
But we can clearly understand the distinction between the apparent ($R_{\rm h}$),
particle ($R_{\rm p}$) and event,
\begin{equation}
R_{\rm e}\equiv a(t)\int_t^\infty {c\,dt^\prime\over a(t^\prime)}\;, \label{eq:horev}
\end{equation}
\newline\noindent
horizons via the use of Equation~(\ref{eq:nullgeo}), and why the proper size of the observable
Universe must be related to $R_{\rm h}$ through Equation~(\ref{eq:maxrad}) rather than $R_{\rm p}$
and $R_{\rm e}$.

Differentiating Equation~(\ref{eq:horpart}) with respect to $t$, one gets the null geodesic
Equation~(\ref{eq:nullgeo}) with a `$+$' sign, which describes the propagation of a photon
{\it away} from the observer. The solution in Equation~(\ref{eq:horpart}) therefore
represents the maximum proper distance a particle traveled away from us during the time
elapsed since the Big Bang. But this is different from the maximum proper distance a photon
traveled in reaching us which, as we have seen, is given by Equation~(\ref{eq:maxrad}).
While $R_{\gamma,\,{\rm max}}$ is bounded by the apparent horizon $R_{\rm h}$, there is no
limit to $R_{\rm p}(t)$, since $\dot{R}_{\rm p}$ is always greater than $c$, so $R_{\rm p}$
easily grows past $R_{\rm h}$, especially at late times in $\Lambda$CDM, when the cosmological
constant dominates the energy budget and the Universe enters a late de Sitter expansion.
We never again see the photons receding from us, reaching proper distances $\sim$~$R_{\rm p}$,
so regions of the Universe that far away are not observable. As noted earlier, null
geodesics must actually {\it reach} us in order for us to see the photons traveling along them,
providing information on their source.

In contrast, the `event' horizon (Eq.~\ref{eq:horev}) is defined to be the largest comoving
distance from which light emitted now, at time $t$, can ever reach us in the asymptotic future.
If we differentiate this equation with respect to $t$, we get Equation~(\ref{eq:nullgeo}) with
a `$-$' sign, representing a photon propagating towards the observer. The physical meaning of
$R_{\rm e}$ is thus similar to that of $R_{\gamma}$, except that the distance in
Equation~(\ref{eq:horev}) represents a horizon for null geodesics that will connect to us in
our future, not today.  This is the reason why the gravitational (or apparent) horizon is
generally not an event horizon yet, though it may turn into one for some equations-of-state
in the cosmic fluid, which influence the evolution of $R_{\rm h}$.

We note, in this regard, that the apparent and event horizons coincide in the metrics of
de~Sitter, Schwarzschild and Kerr specifically because their spacetime curvatures are
independent of time. As we discuss in \S~3.B, this appears to be the reason why the holographic
principle involves $R_{\rm e}$ in those cases, whereas the correct horizon to use for
cosmologies other than de Sitter is $R_{\rm h}$ (or, to be more precise, it should be
$\eta R_{\rm h}(t)$; see \S~3.A above). In other words, it would be correct to say that
the holographic principle should {\it always} be associated with the gravitational (or apparent)
horizon $R_{\rm h}$, rather than $R_{\rm p}$ or $R_{\rm e}$, and it only appears to involve
$R_{\rm e}$ when the apparent and event horizons coincide.

{\acknowledgement
I am very grateful to the anonymous referee for their exceptional review of this
manuscript, and for suggesting several improvements to the presentation.
\endacknowledgement}

%=====================================================
%
%                           BIBLIOGRAPHY
%
%=====================================================

%=====================================================

\begin{thebibliography}{99}
\bibitem{Layzer:1975} D. Layzer, Sci. Am. {\bf 233} (1975) 56
\bibitem{Price:1996} H. Price, {\it Time's Arrow and Archimedes' Point: New Directions
for the Physics of Time}. Oxford University Press, Oxford (1996)
\bibitem{Albert:2000} D. Z. Albert, in {\it Time and Chance}. Harvard University Press,
Cambridge (2000)
\bibitem{Earman:2006} J. Earman, Stud. Hist. Philos. Sci. Part B Stud. Hist. Philos.
Mod. Phys. {\bf 37} (2006) 399
\bibitem{Carroll:2004} S. M. Carroll \& J. Chen, eprint arXiv:hep-th/0410270 (2004)
\bibitem{Patel:2017} V. M. Patel \& C. H. Lineweaver, Entropy {\bf 19} (2017) 411
\bibitem{Zemansky:1997} M. W. Zemansky \& R. H. Dittman, {\it Heat and Thermodynamics}.
McGraw-Hill, New York (1997)
\bibitem{Frautschi:1982} S. Frautschi, Science {\bf 217} (1982) 593
\bibitem{Albrecht:2002} A. Albrecht, in {\it Science and Ultimate Reality: Quantum
Theory, Cosmology and Complexity}. Cambridge University Press, Cambridge (2002)
\bibitem{Penrose:2004} R. Penrose, in {\it Road to Reality: A Complete
Guide to the Laws of the Universe}. Jonathan Cape, London (2004)
\bibitem{Egan:2010} C. A. Egan \& C. H. Lineweaver, Astroph. J. {\bf 710} (2010) 1825
\bibitem{Vilenkin:1982} A. Vilenkin, PLB {\bf 117} (1982) 25
\bibitem{Linde:1984} A. D. Linde, Lett. Nuovo Cim. {\bf 39} (1984) 401
\bibitem{Hartle:1983} J. B.  Hartle \& S. W. Hawking, PRD {\bf 28} (1983) 2960
\bibitem{Melia:2018a} F. Melia, MNRAS {\bf 481} (2018) 4855
\bibitem{Melia:2019a} F. Melia, Annals Phys. {\bf 411} (2019) 167997
\bibitem{Melia:2020a} F. Melia, {\it The Cosmic Spacetime}. Taylor \& Francis, New York (2020)
\bibitem{Kolb:1994} E. W. Kolb \& M. S. Turner, {\it The Early Universe}. Avalon Publishing,
New York (1994)
\bibitem{Bekenstein:1973} J. D. Bekenstein, PRD {\bf 1973} (1973) 2333
\bibitem{Hawking:1983} S. W. Hawking \& D. N. Page, Commun. Math. Phys. {\bf 87} (1983) 577
\bibitem{Zurek:1982} W. H. Zurek, PRL {\bf 49} (1982) 1683
\bibitem{Page:2005} D. N. Page, New J. Phys. {\bf 7} (2005) 203
\bibitem{Binney:2008} J. Binney \& S. Tremaine, {\it Galactic Dynamics}. Princeton
University Press, Princeton (2008)
\bibitem{Wallace:2010} D. Wallace, Br. J. Philos. Sci. {\bf 61} (2010) 513
\bibitem{Guth:1981} A. H. Guth, PRD {\bf 23} (1981) 347
\bibitem{Linde:1982} A. D. Linde, PLB {\bf 108} (1982) 389
\bibitem{Albrecht:1982} A. Albrecht \& P. J. Steinhardt, PRL {\bf 48} (1982) 1220
\bibitem{Davies:1983} P.C.W. Davies, Nature {\bf 301} (1983) 398
\bibitem{Albrecht:2009} A. Albrecht, J. Phys. Conf. Ser. {\bf 174} (2009) 012006
\bibitem{Page:1983} D. N. Page, Nature {\bf 304} (1983) 39
\bibitem{Penrose:1989} R. Penrose, {\it The Emperor's New Mind}. Oxford University Press, 
Oxford (1989)
\bibitem{Planck:2018} Planck Collaboration, N. Aghanim, Y. Akrami et~al., arXiv e-prints, 
arXiv:1807.06209 (2018)
\bibitem{MeliaLopez:2018} F. Melia \& M. L{\'o}pez-Corredoira, Astron. Astrophys. {\bf 610}
(2018) A87
\bibitem{LiuMelia:2020} J. Liu \& F. Melia, Proc. R. Soc. A {\bf 476} (2020) 20200364
\bibitem{Boltzmann:1895} L. Boltzmann, Nature {\bf 1322} (1895) 413
\bibitem{Boltzmann:1897} L. Boltzmann, Annalen der Physik {\bf 296} (1897) 392
\bibitem{Evans:1994} D. J. Evans \& D. J. Searles, PRE {\bf 50} (1994) 1645
\bibitem{Carter:1973} B. Carter, in Proceedings of the Confrontation of Cosmological
Theories with Observational Data; Krakow, Poland; ed. M. S. Longair. Springer Science \&
Business Media, Netherlands (1973)
\bibitem{Barrow:1988} J. D. Barrow, {\it The Anthropic  Cosmological Principle}.
Oxford University Press, Oxford (1988)
\bibitem{Susskind:1993} L. Susskind, L. Thorlacius \& J. Uglum, PRD {\bf 48} (1993) 3743
\bibitem{Hooft:1993} G. 't Hooft, e-print arXiv:gr-qc/9310026 (1993)
\bibitem{Susskind:1995} L. Susskind, J. Math. Phys. {\bf 36} (1995) 6377
\bibitem{Bousso:2002} R. Bousso, Rev. Mod. Phys. {\bf 74} (2002) 825
\bibitem{Bousso:1999} R. Bousso, JHEP {\bf 9907} (1999) 004
\bibitem{Zurek:1985} W. H. Zurek \& K. S. Thorne, PRL {\bf 54} (1985) 2171
\bibitem{Frolov:1993} V. P. Frolov \& D. N. Page, PRL {\bf 71} (1993) 3902
\bibitem{Sorkin:1997} R. D. Sorkin, e-print arXiv:gr-qc/9705006 (1997)
\bibitem{Flanagan:2000} E. E. Flanagan, D. Marolf \& R. M. Wald, PRD {\bf 62} (2000) 084035
\bibitem{Melia:2018b} F. Melia, Am. J. Phys. {\bf 86} (2018) 585
\bibitem{Lineweaver:2008} C. H. Lineweaver \& C. A. Egan, Phys. Life. Rev. {\bf 5} (2008) 225
\bibitem{Kolb:1981} E. W. Kolb \& M. S. Turner, Nature {\bf 294} (1981) 521
\bibitem{Bousso:2007} R. Bousso, R. Harnik, G. D. Kribs \& G. Perez, PRD {\bf 76} (2007) 043513
\bibitem{Frampton:2008} P. H. Frampton \& T. W. Kephart, JCAP {\bf 6} (2008) 8
\bibitem{Frampton:2009} P. H. Frampton, S.D.H. Hsu, T. W. Kephart \& D. Reeb, Class. Q. Grav. 
{\bf 26} (2009) 145005
\bibitem{Collins:1992} W. Collins, PRD {\bf 45} (1992) 495
\bibitem{Hayward:1999} S. A. Hayward, S. Mukohyama \& M. C. Ashworth, PLA {\bf 256} (1999) 347
\bibitem{Hayward:1998} S. A. Hayward, CQG {\bf 15} (1998) 3147
\bibitem{Bak:2000} D. Bak \& S.-J. Rey, CQG {\bf 17} (2000) L83
\bibitem{Bousso:2005} R. Bousso, PRD {\bf 71} (2005) 064024
\bibitem{Nielsen:2009} A. B. Nielsen \& D.-H. Yeom, IJMP-A {\bf 24} (2009) 5261
\bibitem{Davies:1988} P.C.W. Davies, CQG {\bf 5} (1988) 1349
\bibitem{Frolov:2003} A. Frolov \& L. Kofman, JCAP {\bf 0305} (2003) 009
\bibitem{Wang:2006} B. Wang, Y. Gong \& E. Abdalla, PRD {\bf 74} (2006) 083520
\bibitem{Bikwa:2012} O. Bikwa, F. Melia \& A. Shevchuk, MNRAS {\bf 421} (2012) 3356
\bibitem{Frampton:2009a} P. H. Frampton, JCAP {\bf 10} (2009) 16
\bibitem{Penrose:1987} R. Penrose, in {\it Three Hundred Years of Gravitation}, ed. S. W. Hawking
\& W. Israel, pp. 704. Cambridge University Press, Cambridge (1987) p.~17
\bibitem{Hawking:1970} S. W. Hawking, Nature {\bf 248} (1970) 30
\bibitem{Hawking:1975} S. W. Hawking, Comm. Math. Phys. {\bf 43} (1975) 199
\bibitem{Parikh:2000} M. K. Parikh \& F. Wilczek, PRL {\bf 85} (2000) 5042
\bibitem{Kodama:1980} H. Kodama, Prog. Theor. Phys. {\bf 63} (1980) 1217
\bibitem{Misner:1964} C. W. Misner \& D. H. Sharp, PR {\bf 136} (1964) 571
\bibitem{Hernandez:1966} W. C. Hernandez \& C. W. Misner, ApJ {\bf 143} (1966) 452
\bibitem{Melia:2007} F. Melia, MNRAS {\bf 382} (2007) 1917
\bibitem{Einstein:1997} A. Einstein, in {\it The Collected Papers of Albert Einstein, Vol. 6,
1914-1917}, E. Schucking (consultant). Princeton University Press, Princeton (1997)
\bibitem{Padmanabhan:2010} T. Padmanabhan, {\it Gravitation: Foundations and Frontiers}.
Cambridge University Press, Cambridge (2010)
\bibitem{Padmanabhan:2002} T. Padmanabhan, GRG {\bf 34} (2002) 2029
\bibitem{Gibbons:1977} G. W. Gibbons \& S. W. Hawking, PRD {\bf 15} (1977) 2738
\bibitem{Birrell:1982} N. D. Birrell \& P.C.W. Davies, {\it Quantum Fields in Curved
Space}. Cambridge University Press, Cambridge (1982)
\bibitem{Davies:1986} P.C.W. Davies, L. H. Ford \& D. N.  Page, PRD {\bf 34} (1986) 1700
\bibitem{Medved:2002} A.J.M. Medved, PRD {\bf 66} (2002) 124009
\bibitem{Parikh:2002} M. K. Parikh, PLB {\bf 546} (2002) 189
\bibitem{DiCriscienzo:2010} R. Di Criscienzo, S. A. Hayward, M. Nadalini, L. Vanzo \&
S. Zerbini, CQG {\bf 27} (2010) 015006
\bibitem{Hawking:1968} S. W. Hawking, J. Math. Phys. {\bf 9} (1968) 598
\bibitem{Hayward:1994} S. A. Hayward, PRD {\bf 49} (1994) 831
\bibitem{Melia:2009} F. Melia \& M. Abdelqader, IJMP-D {\bf 18} (2009) 1889
\bibitem{Melia:2013a} F. Melia, Class. Q. Grav. {\bf 30} (2013) 155007
\bibitem{Ben-Dov:2007} I. Ben-Dov, PRD {\bf 75} (2007) 064007
\bibitem{Tian:2015} D. W. Tian \& I. Booth, PRD {\bf 92} (2015) 024001
\bibitem{Faraoni:2011} V. Faraoni, PRD {\bf 84} (2011) 024003
\bibitem{Izquierdo:2006} G. Izquierdo \& D. Pavon, PLB {\bf 633} (2006) 420
\bibitem{Dutta:2010} J. Dutta \& S. Chakraborty, GRG {\bf 42} (2010) 1863
\bibitem{Setare:2007} M. R. Setare, JCAP {\bf 2007} (2007) 023
\bibitem{Karami:2010} K. Karami \& S. Ghaffari, PLB {\bf 685} (2010) 115
\bibitem{Abdolmaleki:2014} A. Abdolmaleki, T. Najafi \& K. Karami, PRD {\bf 89} (2014) 104041
\bibitem{Herrera:2014} R. Herrera \& N. Videla, IJMP-D {\bf 23} (2014) 1450071
\bibitem{Melia:2013b} F. Melia, Astron. Astrophys. {\bf 553} (2013) id. A76 
\bibitem{Starobinsky:1980} A. A. Starobinsky, PLB {\bf 91} (1980) 99
\bibitem{Melia:2018c} F. Melia, EPJ-C Lett. {\bf 78} (2018) 739
\bibitem{Martin:2001} J. Martin \& R. H. Brandenberger, PRD {\bf 63} (2001) 123501
\bibitem{Brandenberger:2001} R. H. Brandenberger \& J. Martin, Modern Physics
Letters A {\bf 16} (2001) 999
\bibitem{Brandenberger:2013} R. H. Brandenberger \& J. Martin, CQG {\bf 30} (2013) 113001
\bibitem{Melia:2019b} F. Melia, EPJ-C Letters {\bf 79} (2019) 455
\bibitem{Melia:2020b} F. Melia, Astron. Nach. {\bf 341} (2020) 812
\bibitem{Steinhardt:2016}  C. L. Steinhardt, P. Capak, D. Masters \& J. S. Speagle, 
ApJ {\bf 824} (2016) 1
\bibitem{Melia:2013c} F. Melia, ApJ {\bf 764} (2013) 72
\bibitem{Melia:2018d} F. Melia, A\&A {\bf 615} (2018) A113
\bibitem{Melia:2014} F. Melia, AJ {\bf 147} (2014) 120
\bibitem{Yennapureddy:2019} M. K. Yennapureddy \& F. Melia, EPJ-C {\bf 79} (2019) 571
\bibitem{Melia:2016} F. Melia \& M. Fatuzzo, MNRAS {\bf 456} (2016) 3422
\bibitem{Abbott:1984} L. F. Abbott \& M. B. Wise, Nucl. Phys. {\bf 244} (1984) 541
\bibitem{Lucchin:1985} F. Lucchin \& S. Materrese, PRD {\bf 32} (1985) 1316
\bibitem{Barrow:1987} J. Barrow, PLB {\bf 187} (1987) 12
\bibitem{Liddle:1989} A. R. Liddle, PLB {\bf 220} (1989) 502
\bibitem{Dabrowski:1989} M. P. Dabrowski \& J. Stelmach, AJ {\bf 97} (1989) 978
\bibitem{Zeldovich:1980} Ya. B. Zeldovich, MNRAS {\bf 192} (1980) 663
\bibitem{Vilenkin:1981} A. Vilenkin, PRD {\bf 24} (1981) 2082
\bibitem{Kibble:1976} T.W.B. Kibble, J. Phys. A {\bf 9} (1976) 1387
\bibitem{Avelino:1996} P. Avelino \& R. Caldwell, PRD {\bf 53} (1996) 5339
\bibitem{Gold:1992} T. Gold, Am. J. Phys. {\bf 30} (1992) 403
\bibitem{Busso:1999} R. Bousso, J. High En. Phys. {\bf 7} (1999) 004
\bibitem{Fischler:1998} W. Fischler \& L. Susskind, eprint arXiv:hep-th/9806039 (1998)
\bibitem{Bengtsson:2011} I. Bengtsson \& J.M.M. Senovilla, PRD {\bf 83} (2011) 044012
\bibitem{Faraoni:2015} V. Faraoni, {\it Cosmological and Black Hole Apparent Horizons}.
Springer, New York (2015)
\end{thebibliography}
\end{document}